\documentclass[aps,prd,nofootinbib,preprintnumbers,twocolumn]{revtex4-1}
\usepackage{amsmath,amsfonts,amssymb,graphicx,graphics,color,hyperref}
\usepackage[latin9]{inputenc}
\usepackage{pifont}
\usepackage{natbib}
\usepackage{subfigure,epsfig,epstopdf}
\usepackage{bm}
\newcommand{\be}{\begin{equation}}
\newcommand{\ee}{\end{equation}}
\newcommand{\bea}{\begin{eqnarray}}
\newcommand{\eea}{\end{eqnarray}}

\begin{document}
\title{Enhanced tau neutrino appearance through invisible decay}

\author{Giulia Pagliaroli $^1$}
\email{$^1$giulia.pagliaroli@lngs.infn.it}
\author{Natalia Di Marco $^2$ and  Massimo
  Mannarelli $^2$}
%\affiliation{{$1$}GSSI}
\address{$^1$ Gran Sasso Science Institute, L'Aquila (AQ), Italy}
\address{$^2$ INFN, Laboratori Nazionali del Gran Sasso, Assergi (AQ),
  Italy}
%\ead{giulia.pagliaroli@lngs.infn.it}
%\author{Natalia Di Marco}
%\affiliation{INFN, Laboratori Nazionali del Gran Sasso, Via G. Acitelli, 22, I-67100 Assergi (AQ), Italy}
%\author{Massimo Mannarelli}
%\affiliation{INFN, Laboratori Nazionali del Gran Sasso, Via G. Acitelli, 22, I-67100 Assergi (AQ), Italy}

\begin{abstract}
The decay of neutrino mass eigenstates leads to a change of the conversion and survival probability of neutrino flavor eigenstates.  
Remarkably, we find that the neutrino decay provides an enhancement of the expected tau appearance signal with respect to the standard oscillation scenario for the long-baseline OPERA experiment. The increase of the  $\nu_\mu \rightarrow \nu_\tau$ conversion probability by the decay of one of the mass eigenstates is due to a reduction of  the ``destructive interference" among the different massive neutrino components. Motivated by the recently released results of the OPERA Collaboration showing a number of observed $\nu_\tau$ events larger than expected, we perform a statistical analysis including the invisible decay hypothesis. We obtain a very mild preference for invisible decays, with a best fit value $\tau_3/m_3 \simeq 2.6\times 10^{-13} $ s/eV, and  a constraint at the 90$\%$ confidence level for the neutrino decay lifetime to be $\tau_3/m_3 \gtrsim 1.3\times 10^{-13}$ s/eV. 
\end{abstract}
\maketitle

\section{Introduction} \label{sec:intro}

Nowadays, the three neutrino oscillation picture is established on a rather firm basis. Results from solar, reactor, atmospheric and accelerator experiments provide compelling evidences for the existence of in flight conversions between neutrinos of different flavors caused by nonzero neutrino masses and mixing angles~\cite{Agashe:2014kda}. 

Nevertheless, nonvanishing neutrino masses indicate the possibility that besides oscillating, neutrinos can decay. Historically, the neutrino decay scenario was the first mechanism proposed for explaining the solar neutrino problem~\cite{Bahcall:1972my,Pakvasa:1972gz}. At present, the possibility of neutrino decay is  constrained by many experimental observations. The Standard Model neutrino decays both through radiative and non-radiative processes are well constrained by the high precision measurement of the cosmic microwave background~\cite{Mirizzi:2007jd, Agashe:2014kda}. Processes involving beyond Standard Model (BSM) physics as \be\label{eq:decay} \nu_i \rightarrow \nu+X\,,\ee are much less constrained. 
Here $\nu_i$ ($i=1,2,3$) is a neutrino mass eigenstate with mass $m_{i}$, while $\nu$ and $X$ are particles in the final state. Actually, $\nu$ can correspond to one or more neutrinos and $X$ can correspond to one or more non-observable particles, typically identified as scalar or pseudoscalar fields. The BSM decay in Eq.\eqref{eq:decay} can be classified as i) {\it visible decay}, in which at least one neutrino in the final state is active; ii) {\it invisible decay}, in which all $\nu$ and $X$ are non-observable particles. In this last case, the final state neutrino particles are identified as {\it sterile} neutrinos $\nu_s$.
Focusing on invisible decay, we expect that a beam of (relativistic) $\nu_{i}$-neutrinos having lifetime, $\tau_i$, is depleted due to the invisible neutrino decay by the factor
\be D_i(L,E)=\exp\left( -  \alpha_i \times L/E\right)\, \label{dec}\ee
where  $E$ is the neutrino energy, $L$ is the distance between the source and the detector and 
\be \alpha_i=\frac{m_i}{\tau_i}\, \ee is the decay parameter.
It is evident that, for a given ratio $L/E$, the neutrino decay is only sensitive to decay parameter.

Limits on $\alpha_i$ have been derived by different neutrino sources. For electron antineutrinos, the most favorable combination is provided by the SN1987A ($L=50\,{\rm kpc}$, $E\sim 20 \,{\rm MeV}$): the observation of electron antineutrinos in Kamiokande-II \cite{Hirata:1987hu} and IMB \cite{Bionta:1987qt} yields the lower limit 
$\alpha_1 \sim \alpha_2 \lesssim 10^{-5}$ eV/s \cite{Frieman:1987as, Baerwald:2012kc}. 
Other bounds are less stringent. As a leading example, the strongest model-independent limits on $\nu_2$ non-radiative  decays are obtained from solar neutrinos for which $E\sim 1 \,{\rm MeV}$ and $L = 1.5\times 10^8\,{\rm km}$; in this case $\alpha_2\lesssim 10^{4}$  eV/s~\cite{Joshipura:2002fb,Bandyopadhyay:2002qg}.
%$\tau_2/m_2 \gtrsim 10^{-4} \,{\rm s/eV}$ 
%
For the visible decay modes, a stringent limit $\alpha_2\lesssim 10^{3}$  eV/s
%$\tau_{2}/m_{2} > 1.1\cdot 10^{-3}$ s/eV
is obtained by the non-observation of solar $\overline{\nu}_e$ appearance in Kamland~\cite{Eguchi:2003gg}.
Independent and highly competitive limits can be also obtained by the observation of high-energy cosmic neutrinos as recently provided by the IceCube detector~\cite{Aartsen:2014gkd}. In this case due to the long baseline, the dependence on the lifetime parameters $\tau_i$ disappears and the main information for discriminating the presence of non-radiative neutrino decays is the observed flavor ratios of neutrinos~\cite{Pagliaroli:2015rca}.
The $\nu_3$ lifetime can be bounded by atmospheric and long-baseline neutrino data obtaining~\cite{GonzalezGarcia:2008ru} %\be \tau_{3}/m_{3} \gtrsim 10^{-10} \text{s eV}^{-1} \ee 
\be \alpha_{3} \lesssim 0.3\times 10^{10} \text{eV/s}\, \label{eq:LimiteMaltoni} \ee
at 90$\%$ C.L., whereas the analysis using only long-baseline data provides less stringent limits \cite{Adamson:2010wi,Gomes:2014yua}
$\alpha_{3} \lesssim 0.3 \div 0.5 \times 10^{12} \text{eV/s}$ at 90$\%$ C.L. In both cases, let us stress that the analysis regards how the $\alpha_3$ decay parameter changes the $\nu_\mu \rightarrow \nu_\mu$ survival probability. 

In the present manuscript, we exploit the recent results released by the long-baseline OPERA experiment~\cite{Agafonova:2015jxn} to perform the first investigation, to the best of our knowledge, of $\alpha_3$  using the $\nu_\mu \to \nu_\tau$ appearance channel. We show that the presence of a decay channel increases the   $\nu_\mu \rightarrow \nu_\tau$ conversion probability $P_{\mu \tau}$, for the OPERA values of $L/E$. This differs from the behavior in the typical $L/E$ range values of MINOS and T2K experiments, for which the neutrino decay leads to a decrease of $P_{\mu \tau}$. In the case of the OPERA experiment, the enhancement is produced by the following mechanism: due to the experimental setup, the ``destructive interference" of different mass eigenstates occurs and the decay of one mass eigenstate can partially wash out this interference increasing the $P_{\mu \tau}$ with respect to the pure oscillations case. On the other hand, for the experimental setups of MINOS and T2K, the decay of one mass eigenstate reduces the ``constructive interference" of the different mass eigenstates decreasing $P_{\mu \tau}$ with respect to the standard oscillations case.
The OPERA experiment~\cite{Acquafredda:2009zz,Agafonova:2013dtp}, located at the Gran Sasso Underground Laboratory (LNGS) of INFN, is designed to investigate the $\nu_\tau$ appearance channel on an event-by-event basis by using an artificial beam (Cern Neutrino to Gran Sasso, CNGS \cite{CNGS}) produced at CERN and mainly composed by $\nu_\mu$. It reported a number of observed $\nu_\tau$ events larger than the expected value. This experimental result triggered our interest for checking whether including the decay hypothesis might improve the fit of the experimental value. We find that this is the case, indeed we obtain that $\alpha_3^\text{BF} \simeq 3.8 \times 10^{12}$ eV/s, but the statistical significance is rather poor, less than $1 \sigma$.  

The present paper is organized as follows. In Sec.~\ref{sec2} we discuss the  $\nu_\mu$ to $\nu_\tau$ conversion probability given by the combination of flavor oscillations and the invisible decay of the $\nu_3$ mass eigenstate. We provide a general analysis for any value of  $L/E$,  comparing the behavior of  $P_{\mu \tau}$ for the OPERA experiment with that for  the MINOS and T2K experiments.   In Sec.~\ref{sec3} we use the recent results on the number of observed $\nu_\tau$ events reported by the OPERA Collaboration in~\cite{Agafonova:2015jxn} to derive a best fit value and an upper limit value for $\alpha_3$.  We draw our conclusions in Sec.~\ref{sec4}.

\section{Combining neutrino oscillation and decay: the model}
\label{sec2}
%%%%%%%%%%%%%%%%%%%%%%%%%%%%%%%%%%%%%%%%%%%%%%%%%%%%%%%%outline
%illustro il modello oscillazioni + decay 
%- approssimazione a due sapori, caso gerarchia normale
%- decadono v3 e v2 in uno sterile (+ majorone). Ma OPERA non è sensibile all'alpha del v2
%
%derivo Pmumu --- casi LBL + Atmo come da figura 1
%
%derivo Pmutau --- plot come da figura 2. accento su prima volta uso appearance e discussione dell'aumento della 
%probabilità totale nel caso peculiare di L/E di OPERA. 
%
%limite pure decay e letteratura correlata.
%%%%%%%%%%%%%%%%%%%%%%%%%%%%%%%%%%%%%%%%%%%%%%%%%%%%%%%

Let us assume that active neutrinos are subject to both standard mixing and invisible decays. Therefore, propagating neutrinos mix among flavor eigenstates in an oscillatory time-reversible manner and disappear due to time-irreversible decay. In this hypothesis, the mass eigenstates evolution is given by 
\begin{equation}
\vert\nu_i(t)\rangle=\vert\nu_i(0)\rangle e^{-iE_i t-\frac{\Gamma_i}{2}t }\,,
\label{eq:neutrino_evolution}
\end{equation}
where 
%\begin{equation}
$E_i\simeq p+ m_i^2/(2p)$ and $\Gamma_i=\alpha_i/E_i$. Clearly, massive neutrinos of any flavor can decay by invisible processes. In the two neutrino flavor approximation, which is adequate to describe long-baseline experiments, both $\nu_3$ and $\nu_2$ can decay.
However, bearing in mind that the baseline of the OPERA experiment is $L_0\simeq 730$ km, the mean neutrino energy is $E_0\simeq 17$ GeV and the stringent limits on the value of $\alpha_2$ found both with solar and SN data (see Sec.~\ref{sec:intro}), we can set $\alpha_2=0$ in our considerations. 
Note that this is implicitly assumed in the analysis reported in~\cite{GonzalezGarcia:2008ru,Adamson:2010wi,Gomes:2014yua} where, in a two flavor approximation, only $\nu_3$ is allowed to decay through the process $\nu_3 \to \nu_s +X$ while $\nu_2$ is considered stable. 

Assuming that the flavor eigenstates are obtained by rotation with the standard mixing matrix 
% $\langle\nu_\mu|=\sum_i
%U^*_{i \mu}\langle\nu_i|$ and $|\nu_\tau(t)\rangle=\sum_j U_{\tau j}|\nu_j(t)\rangle$ by
 \begin{equation}
\mathbf{U} =
\left( \begin{array}{cc}
\cos\theta_{23} & \sin\theta_{23} \\
-\sin\theta_{23} &\cos\theta_{23} \\
\end{array} \right)\,,
\end{equation} 
and upon substituting  $\Gamma_i=\delta_{i3}\alpha_i/E_i$ in Eq.~\eqref{eq:neutrino_evolution}, we find, in agreement with~\cite{Gomes:2014yua}, the survival probability 
%\begin{widetext}

\begin{align}
%\begin{split}
P_{\mu\mu}(E,L,\alpha_3)&= \left(\cos^2\theta_{23}+\sin^2\theta_{23} e^{-\frac{\alpha_3}{2E}L} \right)^2 \nonumber\\ &-4\cos^2\theta_{23}\sin^2\theta_{23}e^{-\frac{\alpha_3}{2E}L}\sin^2\left(\frac{\Delta m^2_{23} L}{4E}\right)\,, \label{eq:Pmumu}	
%\end{split}
\end{align}
%\end{widetext}
and the conversion probability
\begin{align}
%\begin{split}
P_{\mu\tau}(E,L,\alpha_3)&=\cos^2\theta_{23}\sin^2\theta_{23}\left(1-e^{-\frac{\alpha_3}{2E}L}\right)^2 \nonumber \\
&+4\cos^2\theta_{23}\sin^2\theta_{23}e^{-\frac{\alpha_3}{2E}L}\sin^2\left(\frac{\Delta
  m^2_{23} L}{4E}\right)\,, \label{eq:Pmutau}	
\end{align}
%\end{equation}
where $\Delta m^2_{23}=m^2_3-m^2_2$ and we replaced $t\rightarrow
L$. These  probabilities depend on the combination of three different effects:
\begin{enumerate}
\item The flavor eigenstates are superposition of the mass eigenstates, leading to the $\theta_{23}$ dependence;
\item The mass eigenstates have different masses, leading to the $\Delta m^2_{23}$ dependence;
\item The mass eigenstate $\nu_3$ can decay,  leading to the $\alpha_{3}$ dependence.
\end{enumerate}

An interesting aspect is that   $P_{\mu\mu} + P_{\mu\tau} = 1 -\sin^2\theta_{23}(1-e^{-\frac{2\alpha_3}{2E}L})$, meaning that the total number of neutrinos is not conserved if $\alpha_3 \neq 0$. Moreover,  the conversion probability is nonzero even for $\Delta m^2_{32} \rightarrow 0$, corresponding to the   pure decay case. First introduced by Barger in~\cite{Barger:1998xk}, the pure decay case is ruled out at more than 3$\sigma$ by SK~\cite{Ashie:2004mr} atmospheric neutrino data and at 7$\sigma$ by the MINOS data analysis~\cite{Adamson:2011ig}.

\begin{figure}[h!]
\includegraphics[width=8.cm]{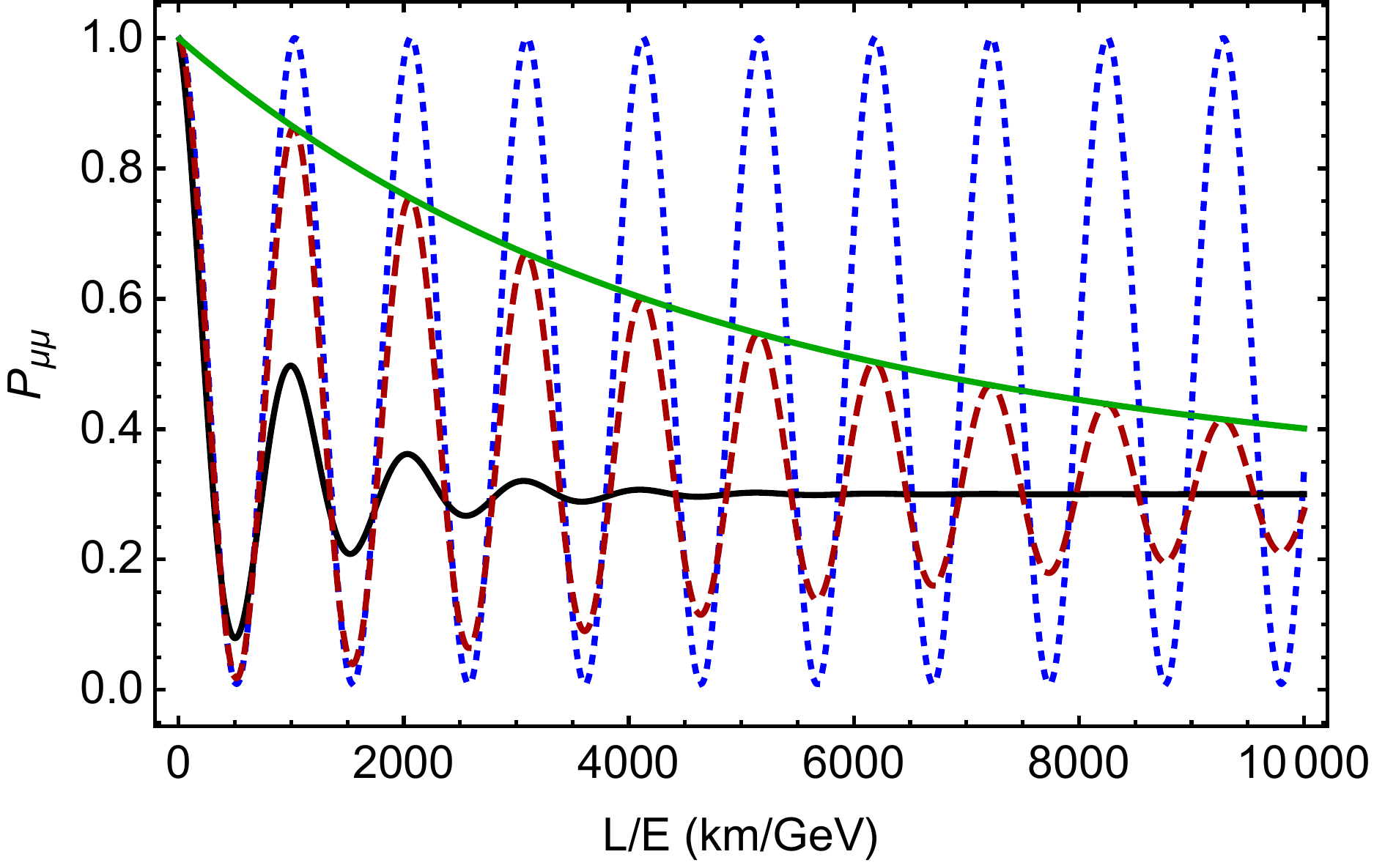}  
\includegraphics[width=8.cm]{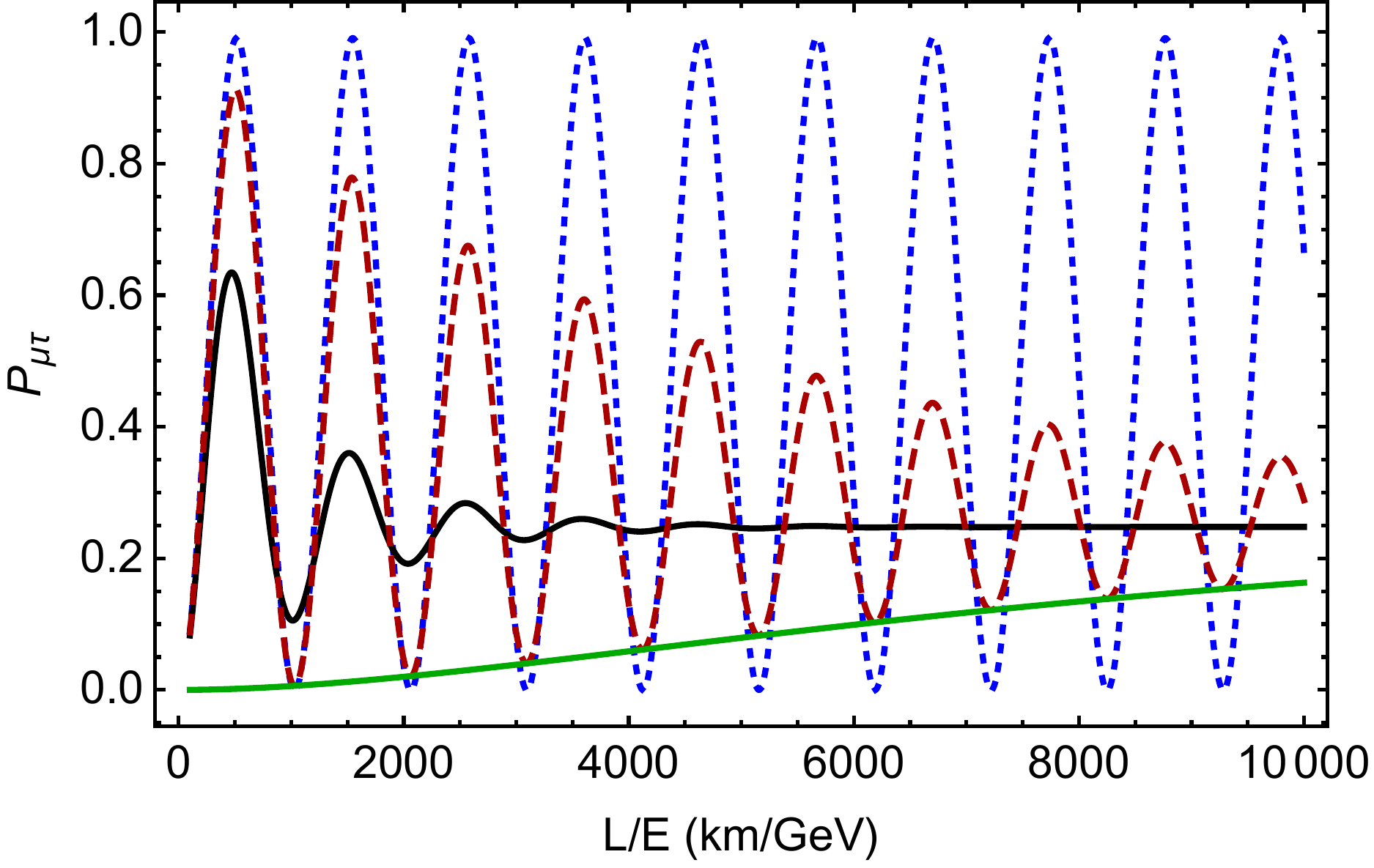}  
\caption{(Color online)  Survival probability (top panel) and  conversion probability (bottom panel),  as a function of the ratio $L/E$ in the two-flavor approximation. The dotted blue lines correspond to the oscillation-only scenario. The red dashed lines refer to the oscillation plus decay hypothesis for $\alpha_3= 10^{11}$ eV/s. The black solid lines correspond to the oscillation plus decay hypothesis for $\alpha_3= 10^{13}$ eV/s. The green solid lines represent the pure decay case with  $\alpha_3= 10^{11}$ eV/s.}
\label{fig:Pmumu}
\end{figure}

The plots in Fig.~\ref{fig:Pmumu} represent the survival probability (top panel) and the conversion probability (bottom panel) as a function of $L/E$ in the two-flavor approximation. In these plots we used  $m_{23}^\text{BF}=\Delta m^2_{23}=2.44\times 10^{-3}$ eV$^2$, $\theta_{23}^\text{BF}=\sin(\theta_{23})^2=0.452$ corresponding to the best-fit values of the global analysis  of~\cite{Gonzalez-Garcia:2014bfa}. The general effect of the neutrino decay is a damping of the standard oscillation amplitude. With increasing  values of  $\alpha_3$, i.e. decreasing  the neutrino lifetime, the  damping effect is stronger. For very short decay times the neutrino oscillations are strongly suppressed even for low values of the ratio $L/E$. For very high values of $L/E$, or equivalently allowing neutrino to propagate through very long distances at a fixed neutrino source energy, the neutrino  conversion probability tends to the constant value $\cos^2\theta_{23}\sin^2\theta_{23}$. The pure decay case is shown in
 Figs.~\ref{fig:Pmumu}  with the  green solid line for $\alpha_3= 10^{11}$ eV/s. 
% is the so-called pure decay case. 
\begin{figure}[th!]
\includegraphics[width=8.cm]{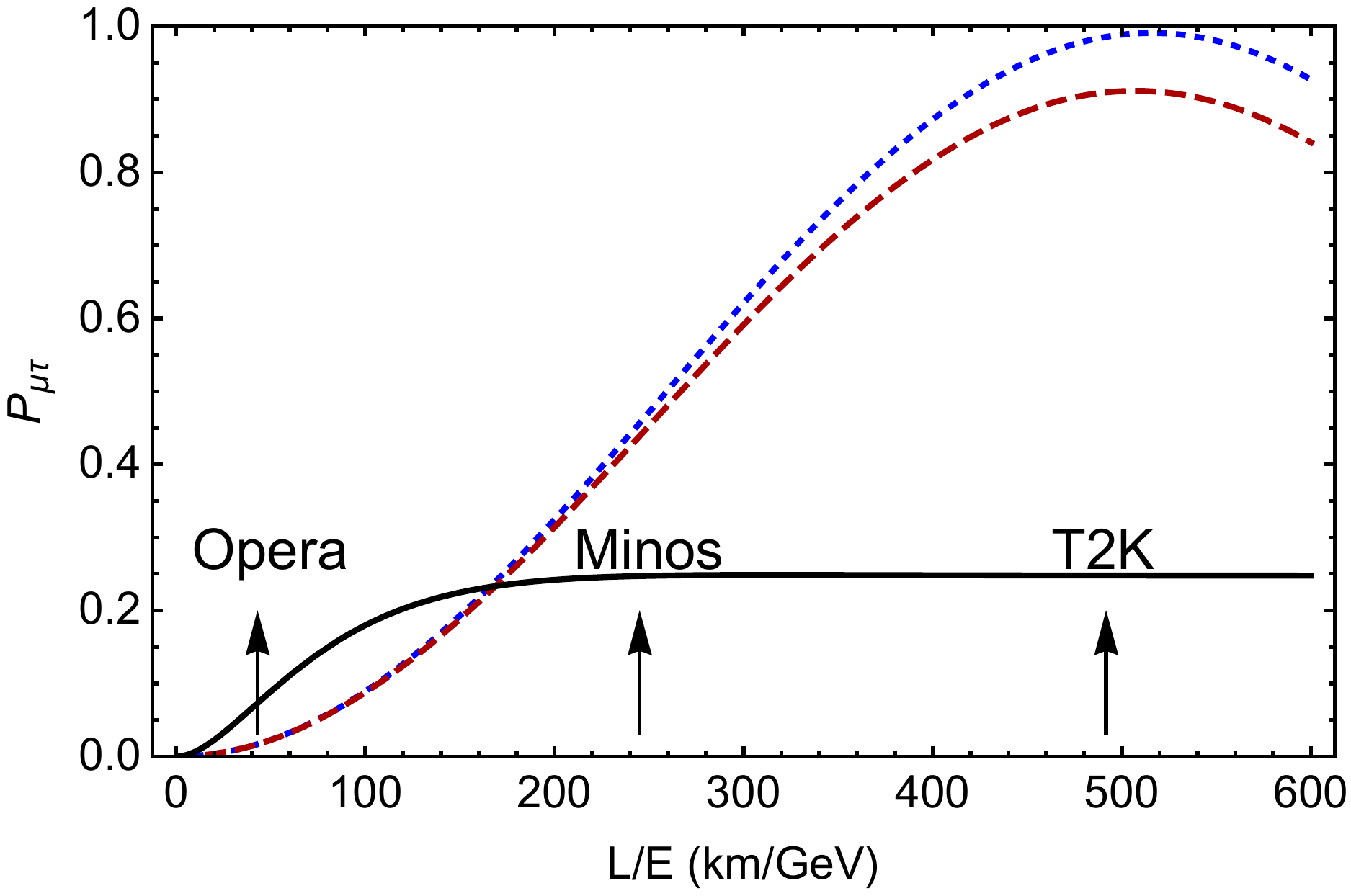}  
\caption{(Color online) Conversion probability   in the region of $L/E$ investigated by OPERA, MINOS and T2K experiments. The dotted blue line corresponds to the oscillation-only scenario. The red dashed line refers to the oscillation plus decay hypothesis for $\alpha_3= 10^{11}$ eV/s. The black solid line refers to the oscillation plus decay hypothesis for $\alpha_3=10^{13}$ eV/s. The arrows indicate the characteristic $L/E$ of OPERA, MINOS, and T2K.}
\label{fig:Pmutau_zoom_LBL}
\end{figure}

As discussed in Sec.~\ref{sec:intro}, the analysis of disappearance data of atmospheric and long baseline experiments, led to the limits on the value of $\alpha_3$ reported in Eq.\eqref{eq:LimiteMaltoni}. The same parameter can be studied with the appearance data of the long baseline OPERA experiment. Fig.\ref{fig:Pmutau_zoom_LBL} shows a zoom of the conversion probability $P_{\mu\tau}$ in  the region of $L/E$ relevant for OPERA, MINOS and T2K. The dotted blue line corresponds to the oscillation-only scenario, the red dashed line refers to the oscillation plus decay hypothesis for $\alpha_3= 10^{11}$ eV/s while the black solid line refers to the oscillation plus decay hypothesis for $\alpha_3=10^{13}$ eV/s. Arrows in the figure indicate the characteristic ratio $\hat{R} = L_0/E_0$ for the different long-baseline experiments, where $L_0$ is the baseline and $E_0$ is the average neutrino beam energy. In particular, $\hat{R}_\text{OPERA}\simeq 730/17$ km$/$GeV, $\hat{R}_\text{MINOS} \simeq 730/3$ km$/$GeV and $\hat{R}_\text{T2K}\simeq 295/2.6$ km$/$GeV. These ratios determine the phase of the transition probability $P_{\mu\tau}$ of Eq.\eqref{eq:Pmutau}. 
When the decay is turned off, i.e. $\alpha_3=0$, the Eq.\eqref{eq:Pmutau} reduces only to the last term due to the interference between the massive components. For the value $\hat{R}_\text{OPERA}$ a situation of destructive interference occurs giving the very small value $P_{\mu\tau}\simeq 0.02$. On the other hand, for $\alpha_3 \rightarrow \infty$, the Eq.\eqref{eq:Pmutau} reduces only to the constant term $\cos^2\theta_{23}\sin^2\theta_{23}$ and $P_{\mu\tau}\simeq 0.25$ (with $\theta_{23}^\text{BF}$).
As a consequence, the OPERA experiment is the only one characterized by an enhancement of the conversion probability when the decay mechanism is turned on with respect to the oscillation-only hypothesis. The opposite situation happens in the case of MINOS and T2K, for which the conversion probability decreases for nonvanishing $\alpha_3$ with respect to the oscillation-only hypothesis. Clearly, an increase of the conversion probability leads to an increase of the number of expected  $\nu_\tau$ events in OPERA. We will discuss this issue in the next section.

\section{Analysis of the OPERA $\nu_\mu \rightarrow \nu_\tau$ appearance results}
\label{sec3}

The OPERA Collaboration recently reported the observation of the 5th candidate $\nu_\tau$ event found in the analysis of an enlarged data sample. The total number of expected events were $2.64 \pm 0.53$ and  $0.25 \pm 0.05$ for signal and background respectively, obtained by assuming $\Delta m^2_{23} = 2.44 \times 10^{-3}$ eV$^2$ and $\sin^2 (2\theta_{23}) = 1$. This result provides a $5.1 \sigma$ evidence for the presence of $\nu_\mu \rightarrow \nu_\tau$ oscillations in the three neutrino flavors framework ~\cite{Agafonova:2015jxn}. 
Given that the number of observed $\nu_\tau$  is larger than expected, and considering the discussion in the previous section,  it seems likely that neutrino decays might be active. 
In the neutrino decay plus oscillation hypothesis, the number of expected $\nu_\tau$ events can be obtained combining the oscillation probability in Eq.~\eqref{eq:Pmutau} with the beam and detector characteristics as follows,
\begin{align} \label{eq:nExpected}
N_{\nu_\tau}^{th}(\alpha_3) &= \epsilon n_{p.o.t} N_{Pb}  \int{ dE \Phi_\mu(E) \sigma_{\nu_\tau}(E)} P_{\mu\tau}(E,L_0,\alpha_3) \,, 	
\end{align}
where $\Phi_\mu(E)$ is the CNGS $\nu_\mu$ flux~\cite{CNGS_fluxes}, $n_{p.o.t}=17.97 \cdot 10^{19}$ is the total number of delivered protons on target (p.o.t) in 5 year of data taking (from 2008 to 2012) \cite{Agafonova:2015jxn}, $\sigma_{\nu_\tau}(E)$ is the $\nu_\tau$ CC cross section \cite{Formaggio:2013kya}, N$_{Pb}=N_A \times M$ is the number of nucleon contained in the 1.2 kton of OPERA lead target \cite{Agafonova:2013dtp}. Finally, the factor $\epsilon$ is the overall experimental $\nu_\tau$ detection efficiency. 

In order to estimate the efficiency factor $\epsilon$, we consider the oscillation only hypothesis, i.e. $\alpha_3=$0, with $\Delta m^2_{23} = 2.44 \times 10^{-3}$ eV$^2$ and $\sin^2 (2\theta_{23})= 1$. Using Eq.(\ref{eq:nExpected}) and setting the efficiency to $1$, the number of expected $\nu_\tau$ events is $\sim 43.5$. Comparing this value with the one quoted by the OPERA Collaboration $2.64$, we assume $\epsilon=2.64/43.5\simeq 6\%$. 
	
\begin{figure}[th!]
\includegraphics[width=8.cm]{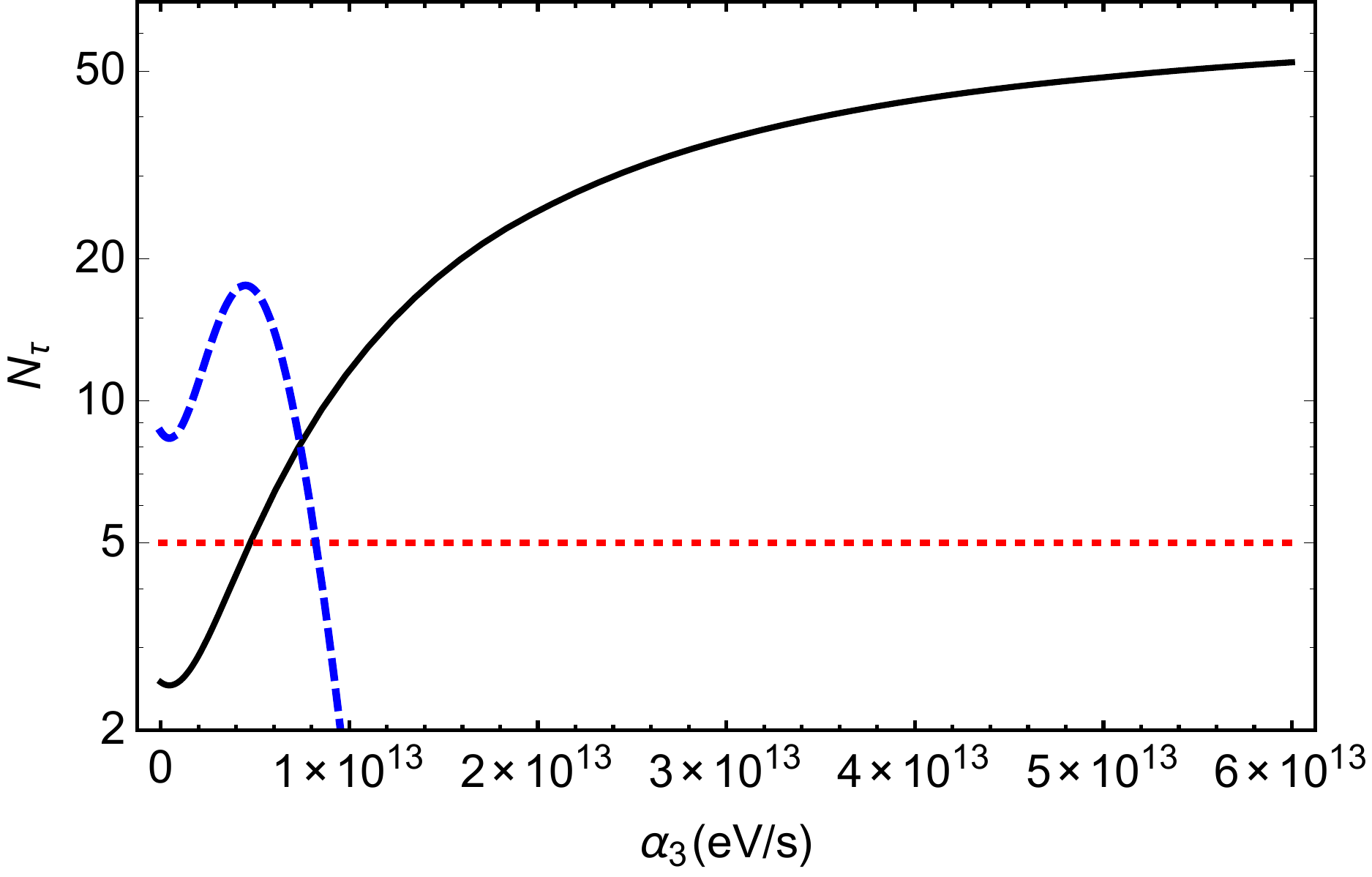}
\caption{(Color online)  Expected number of $\nu_\tau$ events (solid black line) from Eq.~\eqref{eq:nExpected} as a function of the decay parameter $\alpha_3$ and for $m_{23}^\text{BF}$, $\theta_{23}^\text{BF}$~\cite{Gonzalez-Garcia:2014bfa}. The blue dashed line shows, in arbitrary units, the normalized likelihood function. The dotted red line is the number of observed $\nu_\tau$ events in OPERA.}
\label{fig:nev}
\end{figure}	 

The dependence of  $N_{\nu_\tau}^{th}$ on $\alpha_3$  is reported in Fig.~\ref{fig:nev} with a solid black line;  the dotted red line represents the observed number of events in OPERA. The number of expected $\nu_\tau$ events $N_{\nu_\tau}^{th}$ increases as a function of the parameter $\alpha_3$ and saturates to about $67$ events when the decay is complete, i.e. $\alpha_3 \rightarrow \infty$ .  

\begin{figure}[h!]
\includegraphics[width=8.cm]{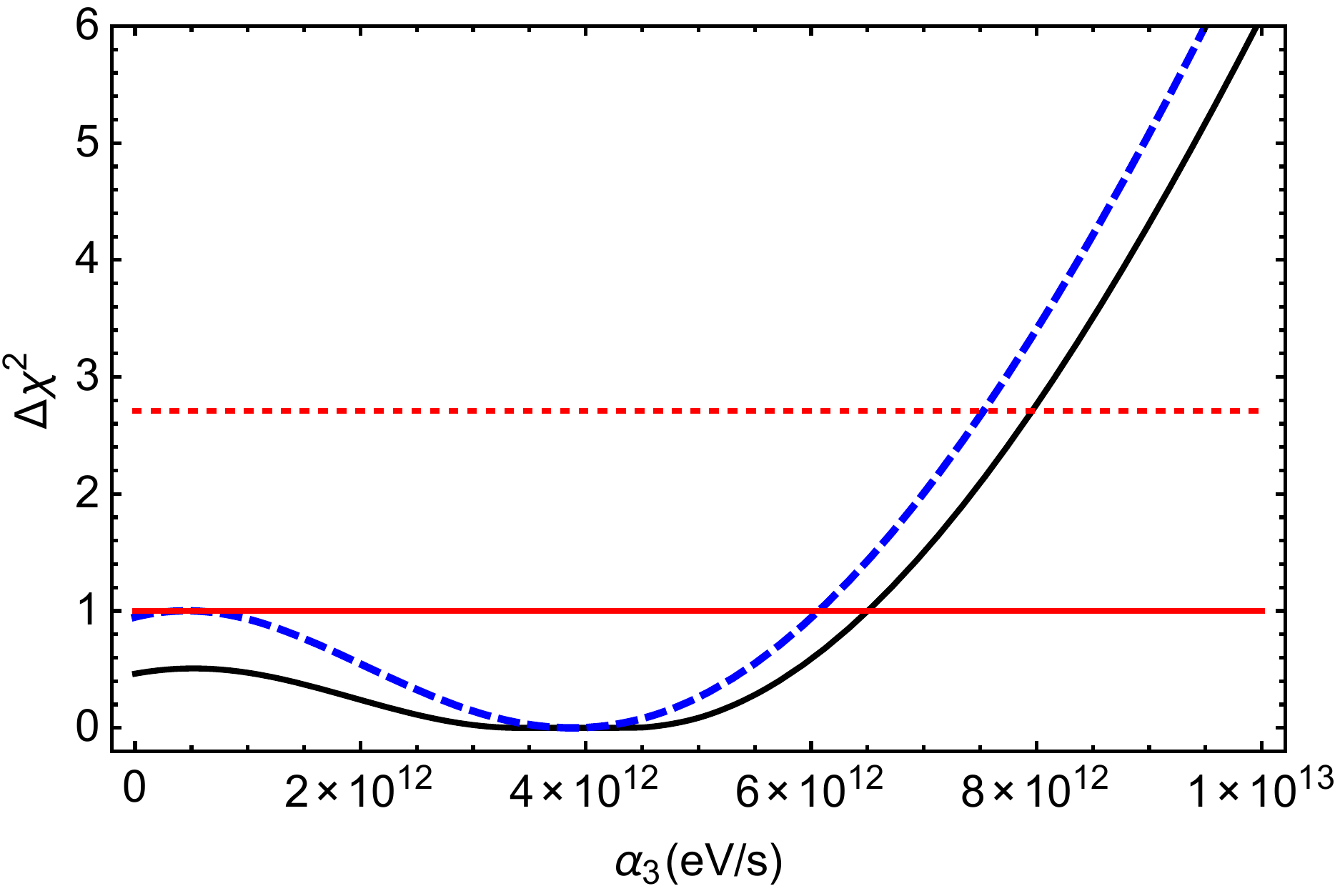}
\caption{(Color online) Value of $\Delta \chi^2$ as a function of $\alpha_3$ considering the others two oscillation parameters fixed to their best-fit values, {\it i.e.} $m_{23}^{BF}$, $\theta_{23}^{BF}$ (blue dashed line) and when the likelihood is marginalized with respect to these two (black solid line). The horizontal lines correspond to the $1\sigma$ (solid red line) and $90\%$ of confidence level (dotted red line).}
\label{fig:lik2flavor}
\end{figure}

The best-fit value for the decay parameter $\alpha_3$ can be estimated by maximizing the Poisson likelihood functions
\be
\mathcal{L}(\alpha_3) \propto \lambda^{n} \times  e^{-\lambda}  
\ee
where $\lambda=N_{\nu_\tau}^{th}(\alpha_3) + b$, $n$ are the observed events, $b=0.25$ are the background events quoted by the OPERA Collaboration. The normalized likelihood function is reported in Fig.~\ref{fig:nev} in arbitrary units with a dashed blue line. In this case the oscillations parameters $\sin^2 \theta_{23}$ and $\Delta m^2_{23}$ are fixed to their best-fit values provided by the global analysis of oscillation data~\cite{Gonzalez-Garcia:2014bfa}. The corresponding $\Delta \chi^2$ function is reported in Fig.~\ref{fig:lik2flavor} with a dashed blue line. We find that OPERA data show a $1\sigma$ preference for the oscillations plus decay model with respect to the oscillation only hypothesis. Indeed the minimum for the $\chi^2$ is characterized by $\alpha_3^\text{BF}\simeq 4.4\times 10^{12} $eV/s. 
To understand the role of the others oscillation parameters we include $\sin^2 \theta_{23}$ and $\Delta m^2_{23}$ as free parameters of the likelihood function. By maximizing the new Poisson likelihood function we find $\alpha_3^\text{BF}\simeq 3.8\times 10^{12} $eV/s, $(\sin^2 \theta_{23})^{BF}=0.458$ and $(\Delta m^2_{23})^{BF}=2.42 \times 10^{-3}$ eV$^2$. In Fig.~\ref{fig:lik2flavor} we show with a black solid line  the $\Delta \chi^2$ function obtained by marginalizing with respect to the other two oscillation parameters, {\it i.e.} allowing them to fluctuate inside their $3\sigma$ ranges of uncertainty\cite{Agashe:2014kda}. The preference for a value of $\alpha_3$ different from zero is stable, however its statistical significance slightly decreases. Using this $\Delta \chi^2$, we can finally set our upper limit at 90 $\%$ of confidence level for the neutrino decay lifetime of $\alpha_3 \lesssim 7.7 \times 10^{12} $ eV/s, or $\tau_3/m_3 \gtrsim 1.3\times 10^{-13}$ s/eV.

\section{Conclusions}
\label{sec4}
Motivated by the recently released OPERA results in the $\nu_\tau$ appearance channel reporting a number of observed events larger than expected, we have performed an analysis of the conversion probability in the presence of neutrino invisible decays. Remarkably, neutrino decay enhances the conversion probability for the OPERA experimental setup, indeed data show a preference for a decay constant different to zero. 
We have demonstrated that for the ratio $L/E$ characteristic of the OPERA experiment the oscillations plus decay model can provide an enhancement of the conversion probability $P_{\mu\tau}$ with respect to the oscillation only hypothesis. This enhancement results in a corresponding increase of the expected number of $\nu_\tau$ CC interactions that better fits the observed number of events (let us remind that the probability of observing $5$ or more candidates with an expectation of $2.64$ signal plus $0.25$ background events is $17\%$ from Poisson statistics \cite{Agafonova:2015jxn}). 
Due to the small statistics, the best fit value we have found for $\alpha_3$ has less than $1 \sigma$ significance and the upper limit at $90 \%$ of confidence level for the neutrino decay lifetime is not competitive with respect to the one already provided by the combined analysis of SK, Minos and T2K, see Eq.~\eqref{eq:LimiteMaltoni}.
However, the information given by the $\nu_\tau$ appearance channel is complementary and could be interesting to strengthen this analysis by including the larger data sets of the  SK detector in this channel~\cite{Abe:2012jj} .

\section*{Acknowledgments}
We would like to thank F.L. Villante for valuable discussions.

\bibliographystyle{apsrev4-1}
\bibliography{BIB_nu}

\end{document}